\documentclass[twocolumn,aps,superscriptaddress,showpacs]{revtex4}
\usepackage{amssymb}
\usepackage{amsmath}
\usepackage{graphicx}
\usepackage[normalem]{ulem}
\usepackage[dvips]{color}

\begin{document}

\title{Why is the nuclear symmetry energy so uncertain at supra-saturation densities?}

\author{Chang Xu}
\affiliation{Department of Physics and Astronomy, Texas A$\&$M
University-Commerce, Commerce, Texas 75429-3011,
USA}\affiliation{Department of Physics, Nanjing University,
Nanjing 210008, China}
\author{Bao-An Li\footnote{Corresponding author, Bao-An\_Li$@$Tamu-Commerce.edu}}
\affiliation{Department of Physics and Astronomy, Texas A$\&$M
University-Commerce, Commerce, Texas 75429-3011, USA}

\begin{abstract}
Within the interacting Fermi gas model for isospin asymmetric
nuclear matter, effects of the in-medium three-body interaction
and the two-body short-range tensor force due to the $\rho$ meson
exchange as well as the short-range nucleon correlation on the
high-density behavior of the nuclear symmetry energy are
demonstrated respectively in a transparent way. Possible physics
origins of the extremely uncertain nuclear symmetry energy at
supra-saturation densities are discussed.
\end{abstract}


\maketitle

\section{Introduction}

The density dependence of nuclear symmetry energy $E_{sym}(\rho)$
is currently a key issue in both nuclear physics and astrophysics,
see e.g.,
Refs.\cite{li3,bro,li2,dan,bar,li1,Sum94,Bom01,Lat04,Ste05a,Lee09}.
Despite much theoretical and experimental effort, our current
knowledge about the $E_{sym}(\rho)$ is still rather poor
especially at supra-saturation densities. Experimentally, some
constraints on the $E_{sym}(\rho)$ at sub-saturation densities
have been obtained recently from analyzing nuclear reaction data,
see, e.g., Refs. \cite{LWC05,tsa,Cen09}. At supra-saturation
densities, however, the situation is much less clear because of
the very limited data available and the few model analysis carried
out so far although some indications of a super-soft
$E_{sym}(\rho)$ at high densities have been obtained from
analyzing the $\pi^+/\pi^-$ ratio in relativistic heavy-ion
collisions\cite{xia}. Theoretically, essentially all available
many-body theories using various interactions have been used in
calculating the $E_{sym}(\rho)$. Unfortunately, the predictions at
supra-saturation densities are very diverse, for a recent review,
see, e.g., Ref. \cite{li1}. Assuming all models are equally
physical and noting that there is no first principle guiding its
high-density limit, it is fair to state that the $E_{sym}(\rho)$
at supra-saturation densities is currently still completely
undetermined. So, why is the $E_{sym}(\rho)$ so uncertain at
supra-saturation densities? This is obviously an important
question that should be addressed timely especially since several
dedicated experiments have now been planned to investigate the
high density behavior of the $E_{sym}(\rho)$ at CSR/China
\cite{CSR}, GSI/Germany \cite{GSI}, MSU/USA \cite{MSU} and
RIKEN/Japan \cite{RIKEN}. Identifying the causes for the uncertain
high-density $E_{sym}(\rho)$ may help experimentalists to decide
what experiments to do and what observables to measure. While we
can not fully answer this question, we identify several important
factors and demonstrate their effects on the high-density behavior
of the $E_{sym}(\rho)$ using probably the simplest many-body
theory available, namely the interacting Fermi gas model for
isospin asymmetric nuclear matter, e.g., Ref. \cite{pre}. There
are many long-standing physical issues on how to treat quantum
many-body problems at high densities, the various techniques used
in different many-body theories may be among the possible origins
of the very uncertain $E_{sym}(\rho)$ at supra-saturation
densities. Nevertheless, it is still very useful to examine
effects of some common ingredients used in most many-body
theories, such as the three-body and tensor forces, in the
simplest model possible. While the interacting Fermi gas model can
not be expected to describe all properties of infinite nuclear
matter and finite nuclei as accurately as those more advanced
microscopic many-body theories, it does give an analytical
expression for the $E_{sym}(\rho)$ in terms of the
isospin-dependent strong nucleon-nucleon (NN) interaction in a
physically very transparent way \cite{xuli2}. Most importantly,
the key underlying physics responsible for the uncertain
$E_{sym}(\rho)$ at supra-saturation densities can be clearly
revealed. In particular, effects of the spin-isospin dependent
effective three-body force, the density dependence of the
in-medium short-range tensor forces and the short-range nucleon
correlation can be demonstrated clearly. The results are expected
to be useful for not only understanding predictions of the various
many-body theories but also ultimately determining the
$E_{sym}(\rho)$ at supra-saturation densities.

\section{Symmetry energy within the interacting Fermi gas model}
According to the well-known Lane potential \cite{Lan62}, the
single-nucleon potential $U_{n/p}$ can be well approximated by
\begin{equation}
U_{n/p}(\rho,k)\approx U_0(\rho,k) \pm U_{sym}(\rho,k)\delta
\end{equation}
where the $U_0(\rho,k)$ and $U_{sym}(\rho,k)$ are, respectively,
the isoscalar and isovector (symmetry) nucleon potentials. Within
the interacting Fermi gas model for isospin asymmetric nuclear
matter \cite{pre}, the nuclear symmetry energy can be explicitly
expressed as (for detailed derivation of this formula, please see
Ref. \cite{xuli2} and references therein).
\begin{eqnarray}\label{Esym}
&&E_{sym}(\rho) =E_{sym}^{kin} + E_{sym}^{pot1} + E_{sym}^{pot2}
\\ \nonumber &=&
\frac{1}{6} \frac{\partial t}{\partial k}\mid _{k_F}\cdot k_F +
\frac{1}{6} \frac{\partial U_0}{\partial k}\mid _{k_F}\cdot k_F +
\frac{1}{2}U_{sym}(\rho,k_F),
\end{eqnarray}
where $t(k)=\hbar ^2 k^2 / 2m$  is the kinetic energy, $m$ is the
average nucleon effective mass and $k_F=(3\pi^2\rho/2)^{1/3}$
(Thomas-Fermi Approximation \cite{TFM}) is the nucleon Fermi
momentum in symmetric nuclear matter at density $\rho$. We notice
here that the Eq.(\ref{Esym}) is identical to the one derived
earlier by Brueckner, Dabrowski and Haensel \cite{bru64,Dab73}
using K-matrices within the Brueckner theory. From
Eq.(\ref{Esym}), it is seen that the symmetry energy
$E_{sym}(\rho)$ is only dependent on the single-particle kinetic
and potential energies at the Fermi momentum $k_F$. The first part
$E_{sym}^{kin}=\frac{\hbar ^2}{6m}
(\frac{3\pi^2}{2})^{\frac{2}{3}} \rho^{\frac{2}{3}}$ is the
trivial kinetic contribution due to the different Fermi momenta of
neutrons and protons; $E_{sym}^{pot1}=\frac{1}{6} \frac{\partial
U_0}{\partial k}\mid _{k_F}\cdot k_F$ is due to the momentum
dependence of the isoscalar potential and also the fact that
neutrons and protons have different Fermi momenta; while the
$E_{sym}^{pot2}= \frac{1}{2}U_{sym}(\rho,k_F)$ is due to the
explicit isospin dependence of the nuclear strong interaction. The
$U_0(\rho_0,k)$ at normal nuclear density $\rho_0$ is relatively
well determined from the nucleon optical potential obtained from
the Dirac phenomenological model analysis of nucleon-nucleus
scattering data \cite{Ham90}. Moreover, interesting information
about the $U_0(\rho,k)$ at abnormal densities in a broad momentum
range has been obtained from transport model analysis of nuclear
collective flow in heavy-ion reactions \cite{Dan00}. For the
momentum dependent part of the isoscalar potential $U_0(k,\rho)$,
we use here the well-known GBD (Gale-Bertsch-Das Gupta)
parameterization \cite{gal}
\begin{equation}
U_{GBD}(\rho,k)=\frac{-75\rho / \rho_0}{1+(k/(\Lambda k_F))^2}
\end{equation}
where $\Lambda=1.5$. The $E_{sym}^{pot1}$ is then given by
\begin{eqnarray}
E_{sym}^{pot1} = 75\rho /
\rho_0\frac{1/(3\Lambda^2)}{(1+(1/\Lambda^2))^2}.
\end{eqnarray}
The GBD potential describes reasonably well the nucleon-nucleus
optical potential and has been widely used in transport model
simulations of heavy-ion reactions \cite{Csernai}. Similar to the
$E_{sym}^{kin}$, the $E_{sym}^{pot1}$ always increases with
density. However, the $U_{sym}(\rho,k)$ is very poorly known
especially at high densities/momenta \cite{li1}. To reveal the
fundamental physics responsible for the uncertain high-density
behavior of the $E_{sym}(\rho)$, we denote $u_{T0}= u'_{np}$ as
the n-p interaction in momentum-space in the isosinglet (T=0)
channel, while $u_{T1}= u_{nn}=u_{pp}=u_{np}$ is the nuclear
strong interaction in the isotriplet (T=1) channel. In the latter,
the charge independence of strong interaction has been assumed.
Then the single-nucleon mean-field potentials are \cite{bom,Zuo05}
\begin{eqnarray}
U_n(\rho,k)= u_{nn} \frac{\rho_n}{\rho} +  u_{np}
\frac{\rho_p}{\rho}
=u_{T1}\frac{\rho_n}{\rho} + u_{T1}\frac{\rho_p}{2\rho}+ u_{T0}\frac{\rho_p}{2\rho}\\
\nonumber U_p(\rho,k) = u_{pp} \frac{\rho_p}{\rho} +  u_{pn}
\frac{\rho_n}{\rho}=u_{T1}\frac{\rho_p}{\rho} +
u_{T1}\frac{\rho_n}{2\rho}+ u_{T0}\frac{\rho_n}{2\rho}.
\end{eqnarray}
Therefore,
\begin{eqnarray}
&&U_0(\rho,k)= \frac{1}{2} (U_n+U_p)= \frac{1}{4} (3u_{T1} +
u_{T0}),\\\nonumber &&U_{sym}(\rho,k)= \frac{1}{2 \delta}
(U_n-U_p) = \frac{1}{4}(u_{T1}-u_{T0}).
\end{eqnarray}
Thus, the $U_{sym}(\rho,k)$ measures the explicit isospin
dependence of the nuclear strong interaction, namely, if the n-p
interactions were the same in the isosinglet and isotriplet
channels, then the $U_{sym}(\rho,k)$ would vanish. Currently, the
calculation of $U_{sym}(\rho,k)$ is rather model dependent
\cite{Das03,Zuo05}. In fact, its value can be either positive or
negative at high densities/momenta \cite{li04x,ria1,ria2}, leading
to the dramatically different predictions on the high-density
behavior of the $E_{sym}(\rho)$. In coordinate space, in terms of
the two-body NN interactions $V_{T0}(r_{ij})$ and $V_{T1}(r_{ij})$
for the isosinglet T=0 and isotriplet T=1 channels, respectively,
we have \cite{pre}
\begin{equation}
E_{sym}^{pot2}=\frac{1}{2} U_{sym}(\rho,k_F)=
\frac{1}{4}(\widetilde{V_{T1}}-\widetilde{V_{T0}}),
\end{equation}
where
\begin{equation}
\widetilde{V_{T0}}=\frac{1}{2}\rho\int V_{T0}(r_{ij})d^3r_{ij},~~
\widetilde{V_{T1}}=\frac{1}{2}\rho\int V_{T1}(r_{ij})d^3r_{ij}.
\end{equation}
It is clear that $E_{sym}^{pot2}$ is determined by the competition
between the $\widetilde{V_{T1}}$ and $\widetilde{V_{T0}}$, and
thus by the isospin dependence of the in-medium NN interactions.
The latter is largely unknown and is actually a major thrust of
research at various radioactive beam facilities around the world.
In fact, how the nuclear medium may modify the bare NN interaction
and properties of hadrons has long been one of the most critical
issues in nuclear physics. For instance, it is very difficult to
obtain the empirical saturation properties of symmetric nuclear
matter by using the bare NN interactions within non-relativistic
many-body theories. Various in-medium effects, such as the
many-body force \cite{vau,zuo}, the relativistic effect relating
the in-medium interactions to free space NN scattering
\cite{broc,fuc}, or the in-medium tensor force due to both the
$\pi$ meson and $\rho$ meson exchanges \cite{bro1,bro2,Rap99},
have to be considered to obtain a reasonably good description of
saturation properties of symmetric nuclear matter. How these
effects manifest themselves in dense neutron-rich matter may
affect the high-density behavior of the nuclear symmetry energy.

\section{Effects of the spin-isospin dependent three-body
force} In this section, we examine effects of the spin-isospin
dependent three-body force on the symmetry energy at
supra-saturation densities. As it has been shown repeatedly in the
literature, see, e.g., Refs. \cite{vau,Oni78,dec,Gra89}, a
zero-range three-body force can be reduced to an effective
two-body force
\begin{eqnarray}
V_{d}=t_0(1+x_0P_{\sigma})\rho^{\alpha}\delta(r),
\end{eqnarray}
where $t_0$, $\alpha$ and $x_0$ are parameters and $P_{\sigma}$ is
the spin-exchange operator. Depending on how many and which
properties of finite nuclei and nuclear matter are used in
reducing the three-body force, various values have been used for
both the $x_0$ and $t_0$ parameters. Moreover, the
density-dependence $\rho^{\alpha}$ is often used to mimic
additional in-medium effects such as the many-body force. For
instance, in relativistic approaches, the dressing of the
in-medium spinors will also introduce a density-dependence to the
interaction, which can lead to similar effects as the many-body
force. The parameter $\alpha$ can thus take on any value between 0
and 1. The parameter $x_0$ controls the relative contributions to
the three-body force from the isosinglet and isotriplet NN
interactions. The three-body term has been included in many
effective interactions, such as the Skyrme and Gogny interactions,
in both phenomenological, e.g., Refs. \cite{Cha97,Sto03} and
microscopic, see, e.g., Ref. \cite{Don09}, many-body theories for
isospin asymmetric nuclear matter. Noticing that the potential
energies due to the three-body force in the T=1 and T=0 channels
are, respectively (for details, see Table II and the corresponding
equation $E^{ST}$ in Ref.\cite{dec}),
\begin{eqnarray}
V_d^{T1}=\frac{1-x_0}{2} \frac{3t_0}{8} \rho ^{\alpha +1}, \,\,\,
V_d^{T0}=\frac{1+x_0}{2} \frac{3t_0}{8} \rho ^{\alpha +1},
\end{eqnarray}
one sees immediately that the terms containing $x_0$ cancel out in
calculating the EOS of symmetric nuclear matter. In fact, among
the normally 13 parameters used in the Hartree-Fock calculations
with the Gogny force $x_0$ is the only one having this special
feature. To evaluate quantitatively the symmetry energy we use the
Gogny central force \cite{dec}
\begin{equation}
V_c(r)=\sum_{i=1,2}(W_i+B_iP_{\sigma}-H_iP_{\tau}-M_iP_{\sigma}P_{\tau})_ie^{-r^2/\mu_i^2},
\end{equation}
where $P_{\tau}$ is the spin exchange operator and the values of
the parameters $W$, $B$, $H$, $M$ and $\mu$ are taken directly
from Ref.\cite{dec}. It is necessary to stress here that since we
are aiming at a qualitative understanding of why the symmetry
energy is so uncertain at supra-saturation densities, the model
parameters are not re-tuned self-consistently to reproduce any
existing constrain on the symmetry energy at and/or below the
normal nuclear matter density $\rho_0$. The resulting
$E_{sym}^{pot2}$ is
\begin{equation}
E_{sym}^{pot2}=
-\sum_{i=1,2}(\frac{H_i}{4}+\frac{M_i}{8})\pi^{\frac{3}{2}}\mu_i^3
\rho -(1+2x_0)\frac{t_0}{8}\rho ^{\alpha +1}.
\end{equation}
\begin{figure}[htb]
\centering
\includegraphics[width=9.5cm]{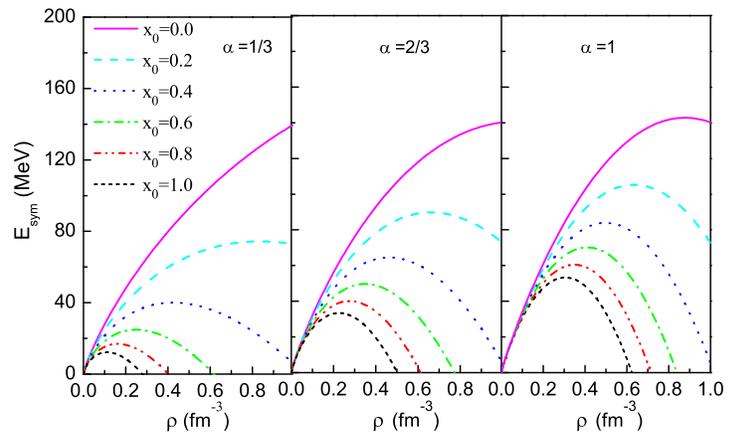}
\caption{(Color online) The symmetry energy obtained with
different spin and density dependences in the three-body force.}
\end{figure}
Shown in Fig.1 are the symmetry energy functions obtained with
different values for the $x_0$ and $\alpha$ parameters. It is
clearly seen that for a given value of the $\alpha$ parameter, the
$x_0$ controls the high density behavior of the symmetry energy.
By varying the $x_0$ one can easily cover the whole range of
symmetry energy calculated within the Hartree-Fock approach using
over one hundred Skyrme and Gogny forces \cite{Sto03} without
changing anything in the EOS of symmetric nuclear matter. In
particular, with $x_0=1$ and $\alpha=1/3$ as in the original Gogny
force \cite{dec}, then only the (S=1, T=0) spin-isospin n-p
interaction contributes to the three-body force and the symmetry
energy, the $E_{sym}(\rho)$ drops quickly to zero above certain
supra-saturation densities. We regard this kind of $E_{sym}(\rho)$
as being super-soft. Overall, depending on the value of the
parameter $x_0$, the symmetry energy can be either stiff that
keeps increasing with density or becomes super-soft above certain
supra-saturation densities.

In the 2003 survey by J. R. Stone et al. \cite{Sto03} of 87 Skyrme
interactions, the $x_0$ ($x_3$ in the notation of Ref. \cite{Sto03})
ranges between $-1.56$ to $1.92$. As a special example, we notice
that B. A. Brown used $x_0$ between 0.03 to 0.9 with the SKX
\cite{Brown98}. The values of $x_0$ used in Fig.1 are thus within
the uncertainty range of $x_0$ found in the literature. Moreover, we
notice that the result here is consistent with the earlier finding
that the EOS of pure neutron matter is very sensitive to the $x_0$
parameter \cite{Brown98,Pethick}. To our best knowledge, it is
currently not clear how to fix the $x_0$ alone experimentally.
Nevertheless, it is worth noting that the cross section of $pn$
charge exchange reactions and the symmetry potential $U_{sym}$
extracted from the isospin dependence of the nucleon optical
potentials are all directly related to the spin-isospin dependent
nuclear interaction. They may thus be used to constrain the value of
$x_0$. Moreover, the density dependence of the nuclear symmetry
energy itself can be used to constrain the $x_0$ should it be
determined experimentally in the future. However, as we shall
discuss in the next section, the in-medium tensor force may have
similar effects on these observables.\\

\section{ Effects of the in-medium short-range tensor force and
nucleon correlation} Studies based on microscopic many-body
theories indicate consistently that the symmetry energy is
dominated by the isosinglet (S=1,T=0) channel \cite{bom,Die03}.
While measured properties of deuterons indicate unambiguously that
a tensor force is at work in the (S=1,T=0) channel. It is also
well known that the $\pi$ and $\rho$ meson exchanges contribute to
the intermediate-range attractive and the short-range repulsive
tensor forces, respectively, according to \cite{bro1}
\begin{equation}
V_T^{\pi}(r)=\frac{f^2_{N_{\pi}} m_{\pi}}{4\pi} \tau_1 \cdot
\tau_2 ( -S_{12})[\frac{e^{-m_{\pi}r}}{(m_{\pi}r)^3} +
\frac{e^{-m_{\pi}r}}{(m_{\pi}r)^2} +
\frac{e^{-m_{\pi}r}}{3m_{\pi}r}],\nonumber
\end{equation}
and
\begin{equation}
V_T^{\rho}(r)=\frac{f^2_{N_{\rho}} m_{\rho}}{4\pi} \tau_1 \cdot
\tau_2 (S_{12})[\frac{e^{-m_{\rho}r}}{(m_{\rho}r)^3} +
\frac{e^{-m_{\rho}r}}{(m_{\rho}r)^2} +
\frac{e^{-m_{\rho}r}}{3m_{\rho}r}],
\end{equation}
where $f^2_{N_{\pi}}/4\pi=0.08$ and $f^2_{N_{\rho}}/m^2_{\rho}
\simeq 2f^2_{N_{\pi}}/m^2_{\pi}$. The $S_{12}=6(\vec{S}\cdot
\vec{r})^2/r^2-2\vec{S}^2=4S^2P_2(cos(\theta))$ is the tensor
operator. We notice that some effective interactions, such as the
Paris force \cite{paris}, only consider the tensor force due to the
$\pi$ exchange and has a short-range cut-off. As shown in Fig.3 of
Ref.\cite{Otsuka05}, the short range behavior of the tensor forces
used in popular effective interactions differ dramatically.
Moreover, different short range cut-offs are normally introduced,
e.g., in studying the single particle energy levels in rare isotopes
\cite{Otsuka05,Otsuka06}. Unfortunately, the different behaviors of
the tensor force at short distance being cut-out, thus not probed
from studying the energy levels of single particles, affect
dramatically the high density behavior of the symmetry energy.
Furthermore, in-medium properties of the $\rho$ meson may affect the
strength of the short-range tensor force. Because the symmetry
energy is very sensitive to the competition between the isosinglet
($\widetilde{V_{T0}}$) and isotriplet ($\widetilde{V_{T1}}$)
channels, the contribution of the in-medium short-range tensor force
that exists only in the T=0 channel may thus affect significantly
the high-density behavior of the $E_{sym}(\rho)$. To investigate
effects of the in-medium tensor force,  we use the Brown-Rho Scaling
(BRS) for the in-medium $\rho$ meson mass according to $
m_{\rho}^{\star}/m_{\rho} = 1-\alpha_{BR}\cdot \rho/\rho_0 $
\cite{bro2}. Shown in Fig.2 is the radial part of the total tensor
force $V_T=V_T^{\rho}(r)+V_T^{\pi}(r)$ at $\rho=0, \rho_0, 2\rho_0,
\,$ and $3\rho_0$, respectively, with the BRS parameter
$\alpha_{BR}=0.2$. As one expects, the total tensor force becomes
more repulsive in denser matter when the $\rho$ meson mass is
reduced according to the BRS. While the experimental evidence for
the BRS is still not very clear, we use it here as an effective way
of adjusting the in-medium strength of the tensor force. This is
useful to mimic different ways of involving the tensor force in
many-body calculations in the literature.

\begin{figure}[htb]
\centering
\includegraphics[width=9cm]{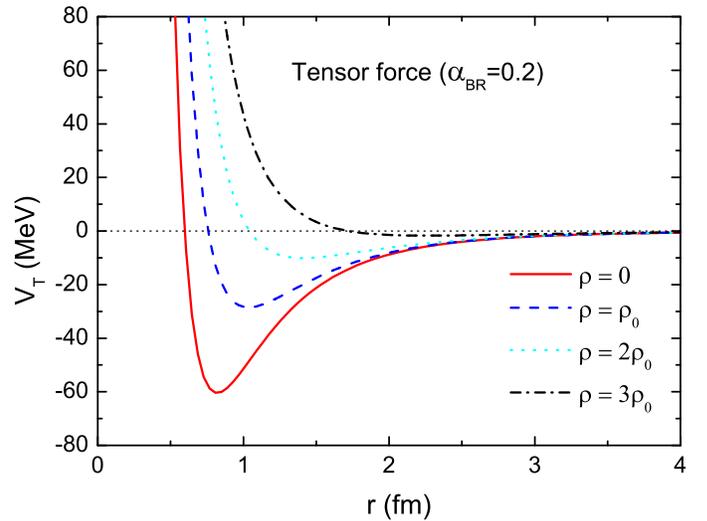}
\caption{(Color online) The radial part of the tensor force
$V_T=V_T^{\rho}(r)+V_T^{\pi}(r)$ at $\rho=0, \rho_0, 2\rho_0,
\,$and$\, 3\rho_0$ with the BRS parameter $\alpha_{BR}=0.2$.}
\end{figure}
\begin{figure}[htb]
\centering
\includegraphics[width=9cm]{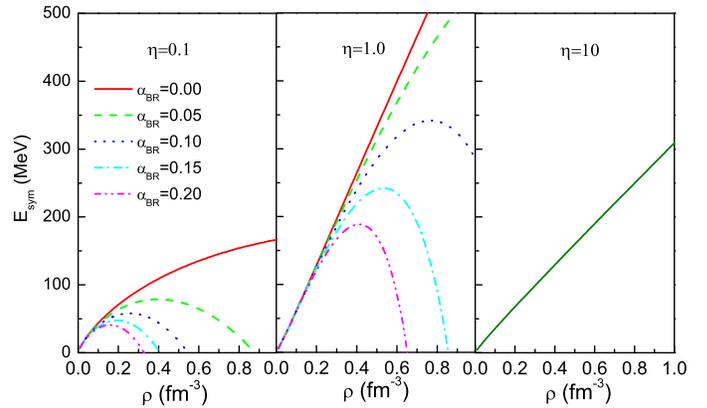}
\caption{(Color online) The symmetry energy with different values
of the BRS parameter $\alpha_{BR}=0, 0.05, 0.10, 0.15, 0.20$ using
three values for the correlation parameter $\eta$.}
\end{figure}

Noticing that the tensor force gives no contribution at the
mean-field level to the potential energy with spherically symmetry
tensor correlations \cite{ris}, we assume here that the tensor
force acting in the isosinglet n-p channels in nuclear matter
behaves the similar way as in deuterons, namely, the tensor
operator $S_{12}$ is a constant of 2. Moreover, we introduce a
two-step tensor correlation function, i.e. $f(r)=0$, for $r< r_c $
and $f(r)=1$, for $r \geq r_c$ where $r_c =\eta
(3/4\pi\rho)^{1/3}$ is the ``healing distance" or short-range
cut-off. Thus the tensor contribution to the isospin T=0 channel
symmetry energy is $\widetilde{V_{T}}=\int f(r)
[V_T^{\rho}(r_{ij})+V_T^{\pi}(r_{ij})]d^3r_{ij}$. The parameter
$\eta$ is used to effectively vary the short range cut-off. The
density dependence in the $r_c$ reflects the design that the size
of nucleons shrinks as the density increase. With $\eta=2$, the
two nucleons will always keep in touch on their surfaces while
their sizes decrease with increasing density.

Shown in Fig.3 are the total symmetry energy functions obtained
using three typical values for the $\eta$ parameter. With
$\eta=10$, the $r_c$ is so large that the tensor contribution to
the $E_{sym}^{pot2}$ is completely cut off. The $E_{sym}(\rho)$
thus keeps increasing with density due to the $E_{sym}^{kin}$ and
the $E_{sym}^{pot1}$ terms. With smaller $\eta$ values, the high
density behavior of the $E_{sym}(\rho)$ is controlled by the BRS
parameter $\alpha_{BR}$. Similar to varying the $x_0$ parameter in
the three-body force, by varying the $\alpha_{BR}$ parameter the
tensor force can lead to $E_{sym}(\rho)$ from stiff to super-soft.
A recent study indicates that a value of $\alpha_{BR}\approx 0.15$
is required to reproduce the measured lifetime of $^{14}C$
\cite{Hol08}. With such an $\alpha_{BR}$, the symmetry energy can
easily become super-soft with $\eta$ between 0.1 and 1. Results
for the three typical cases shown in Fig.3 clearly indicate that
the high-density behavior of the symmetry energy is sensitive to
both the short-range in-medium tensor force and the NN correlation
function. The short-range repulsion generated by the $\rho$-meson
exchange plays the key role in determining the symmetry energy at
supra-saturation densities. In fact, the relationship between the
short-range repulsive tensor force in the isosinglet n-p channel
and the appearance of the super-soft symmetry energy was first
noticed by Pandharipande et al. within variational many-body (VMB)
theories \cite{pan,wir}. Ultimately, because of the dominance of
the repulsive n-p interaction in isospin symmetric nuclear matter
at high densities, it is possible that pure neutron matter is
energetically favored, leading to the negative symmetry energy at
high densities. Of course, this can only occur if the $\rho$
tensor contribution is sufficiently strong due to, for example,
its reduced mass in dense medium. It was also pointed out that the
fundamental reason for the completely different high-density
behaviors of the $E_{sym}$ predicted by the VMB and the
Relativistic Mean Field (RMF) models is the lack of the $\rho$
tensor contribution to the energy in the RMF models \cite{kut}.

To this end, it is interesting to note that the three-body force and
the tensor force can affect similarly the high-density behavior of
the symmetry energy. This is very similar to the situation in
describing the saturation properties of symmetric nuclear matter. It
has been shown that the saturation properties can be equally well
described by either including the three-body force or the in-medium
NN interactions based on the BRS \cite{Rap99,Don09}. Nevertheless,
as it was pointed out in Ref. \cite{Gra89}, since the three-body
force is essentially a convolution of two-body forces, a consistent
three-body force should also include a tensor component. We have
only studied here separately effects of the three-body force and the
two-body tensor force on the high density behavior of the symmetry
energy. Effects including both the two and three-body tensor forces
simultaneously will be investigated in a forthcoming work. In
particular, by varying parameters controlling both the three-body
force and the tensor force together we shall demonstrate how large
the parameter space is in which the symmetry energy may become
negative at supra-saturation densities.\\

\section{Summary} In summary, the high density behavior of the
symmetry energy $E_{sym}$ has long been regarded as the most
uncertain property of dense neutron-rich nuclear matter because
even its trend is still controversial. Within the interacting
Fermi gas model, the $E_{sym}$ can be expressed in terms of the
isospin-dependent NN strong interaction. The high-density behavior
of the $E_{sym}$ is determined by the competition between the
in-medium isosinglet and isotriplet nucleon-nucleon interactions.
Respective effects of the in-medium three-body interaction and the
short-range tensor force on the high-density behavior of the
$E_{sym}$ are examined separately. It is found that the strength
of the spin (isospin) dependence of the three-body force and the
in-medium $\rho$ meson mass in the short-range tensor force are
the key parameters controlling the high-density behavior of the
$E_{sym}$. These findings are useful for understanding why the
nuclear symmetry energy is very uncertain at supra-saturation
densities.

We would like to thank Lie-Wen Chen, Wei-Zhou Jiang, Che Ming Ko,
Hyun Kyu Lee, Rupert Machleidt, Mannque Rho and Wei Zuo for
helpful discussions. One of us (B. A. Li) would also like to thank
the kind hospitality he received at the World Class University
program at Hanyang University in Seoul, Korea where he benefited
from the stimulating discussions on the main issues studied in
this work. This work is supported by the US National Science
Foundation Awards PHY-0652548 and PHY-0757839, the Research
Corporation under Award No.7123 and the Texas Coordinating Board
of Higher Education Award No.003565-0004-2007, the National
Natural Science Foundation of China (Grants 10735010, 10775068,
and 10805026) and by the Research Fund of Doctoral Point (RFDP),
No. 20070284016.

\end{document}